\documentclass[fleqn,usenatbib]{mnras}

\usepackage{newtxtext,newtxmath}
 
\usepackage[T1]{fontenc}
\usepackage{ae,aecompl}

\usepackage{graphicx}	
\usepackage{amsmath}	
\usepackage{amssymb}	

\newcommand{\src}{MAXI~J1820+070}

\newcommand{\nicer}{\emph{NICER}}
\newcommand{\nus}{\emph{NuSTAR}}

\newcommand{\rg}{R$_{\rm{g}}$}

\title[Spin of \src]{ A timing-based estimate of the spin of the black hole in \src}

\author[Bhargava et al.]{
Yash Bhargava,$^{1}$\thanks{E-mail: yash@iucaa.in}
Tomaso Belloni,$^{2}$
Dipankar Bhattacharya,$^{1}$ 
Sara Motta,$^{2}$
and Gabriele Ponti.$^{2,3}$
\\
$^{1}$Inter-University Centre for Astronomy and Astrophysics Post Box No. 4, Ganeshkhind, Pune-411007, India\\
$^{2}$INAF, Osservatorio Astronomico di Brera, Via E. Bianchi 46, I-23807 Merate (LC), Italy\\
$^{3}$Max-Planck-Institut f\"ur Extraterrestrische Physik, Giessenbachstrasse, D-85748, Garching, Germany
}

\date{Accepted XXX. Received YYY; in original form ZZZ}

\pubyear{2021}

\begin{document}
\label{firstpage}
\pagerange{\pageref{firstpage}--\pageref{lastpage}}
\maketitle

\begin{abstract}
\src\ (ASSASN-18ey) is a Black hole X-ray binary discovered in 2018. The brightness of the source triggered multi-wavelength campaigns of this source from different observatories. We analyse the Power Density Spectra obtained from \nicer\ high cadence observations of the source in the hard state. We obtain the evolution of the characteristic frequencies by modelling the PDS. We interpret the characteristic frequencies of various PDS components (both QPOs and broad band noise components) as variability occurring at a particular radius, and explain them in the context of the Relativistic Precession Model. We estimate the dimensionless spin of the black hole at $0.799^{+0.016}_{-0.015}$ by fitting the Relativistic Precession Model. 
\end{abstract}




\section{Introduction}
Black hole binaries (BHBs) are observed to be variable at different time scales \citep{remillard2006ARA&A..44...49R, Done2007A&ARv..15....1D}. Transient BHBs, which are by far the most commonly BHBs observed, typically undergo an hysteresis cycle through a sequence of states characterised by the differences in the spectral and temporal properties \citep{belloni2005A&A...440..207B, remillard2006ARA&A..44...49R, Done2007A&ARv..15....1D}  
These states occupy different positions on the Hardness Intensity Diagram (HID) and in most cases form a hysteresis loop as the source proceeds with an outburst. In particular, the low hard state (LHS) and hard-intermediate state (HIMS) show high variability \citep[$\sim$30\%; ][]{belloni2016ASSL..440...61B}. The power density spectra (PDS) in these states sometimes shows narrow features which are called quasi periodic oscillations \citep[QPOs, see][for a comprehensive review]{Ingram2020arXiv200108758I}. The QPOs observed in the HIMS are often accompanied with broadband low frequency noise \citep{belloni2016ASSL..440...61B}, lie within 0.1-30~Hz, and are classified as type-C QPOs \citep{PBK1999ApJ...520..262P, casella2005ApJ...629..403C}. Soft-intermediate state (SIMS) shows transient QPOs of type-B and High soft state (HSS) shows type-A which differ in the quality factor ($\sim6$ and $\sim3$ respectively) with a fractional r.m.s of 2--4\% \citep[][]{casella2005ApJ...629..403C}. Some  BHBs also show high frequency QPOs \citep[few 100~Hz, HFQPOs,][]{strohmayer2001ApJ...552L..49S, Motta2014MNRAS.437.2554M, Motta2014MNRAS.439L..65M}. In addition to the observed QPOs, the power spectrum of BHBs has broad noise features at higher frequencies \citep{PBK1999ApJ...520..262P, Belloni2002ApJ...572..392B, Motta2014MNRAS.437.2554M, Motta2014MNRAS.439L..65M}. The correlation between the frequencies of different features observed in the PDS is also called Psaltis-Belloni-van der Klis correlation \citep[PBK correlation,][]{PBK1999ApJ...520..262P}. The broad frequencies typically peak at a few 10~Hz and are claimed to be low-frequency counterparts of HFQPOs \citep{PBK1999ApJ...520..262P}.



The measurement of the spin in a black hole (BH) is one of the important challenges in X-ray astronomy. Estimates of the spin can be made using spectroscopic methods in which either  the relativistic broadening of the Fe K$\alpha$ line is measured or the continuum X-ray emission is modelled with thermal components \citep{Miller2009ApJ...697..900M, SpinRev2020arXiv201108948R}. In both cases the innermost radius of the disk is measured and is assumed identical to the innermost stable circular orbit (ISCO). To determine the inner radius of the disk from the thermal continuum modelling, the mass of the BH and the inclination of the disk and the distance to the binary system have to be assumed or measured using alternative methods \citep[see ][ for more details]{SpinRev2020arXiv201108948R}. The detection of HFQPOs provided an alternative method to probe the spin. GRO~J1655$-$40 showed a pair of HFQPOs at 300 and 450~Hz which lead to a lower limit of spin of 0.15 \citep{strohmayer2001ApJ...552L..49S}. \citet{abramowicz2001A&A...374L..19A} interpreted the pair of HFQPOs in GRO~J1655$-$40 as resonances in the orbital and the epicyclic motions. By using the mass measurement from the optical data, the authors constrain the spin of the source in the range of 0.2--0.67. \citet{Nowak1997ApJ...477L..91N} associated the HFQPOs to \textit{g}-modes in the accretion disk for GRS~1915+105. \citet{reynolds2009ApJ...692..869R} explore the association of the HFQPOs to the pressure-driven accretion modes while \citet{mckinney2012MNRAS.423.3083M} link the HFQPOs to the base of a Blandford-Znajek jet. With these associations, it is possible to derive the spin of the source from the HFQPOs observation.

The Relativistic Precession Model \citep[RPM;][]{Stella1998ApJ...492L..59S, Stella1999ApJ...524L..63S}, associates the observed QPOs to different frequencies arising at a particular radius around a BH. In the model, the motion of a test particle is considered around the compact object in a tilted elliptical orbit. The low-frequency type-C QPO is assumed to arise due to the nodal (Lense-Thirring, LT) precession  of the test particle orbit while the lower HFQPO is assumed to arise from the periastron precession. The upper HFQPO is associated with the orbital frequency. The radius at which these frequencies are arising need not be the ISCO but can correspond to a transition region (e.g. inner truncation radius).  
Assuming the RPM, the mass and spin of the BH have been estimated for different BHBs. In an observation of a BHB GRO~J1655$-$40 from the Rossi X-ray Timing Explorer (\emph{RXTE}) archive, a type-C QPO and two high frequency QPOs (HFQPOs) were observed simultaneously \citep[][]{strohmayer2001ApJ...552L..49S,  Motta2014MNRAS.437.2554M}. The detected QPOs were used to determine a mass and spin estimate of the BH consistent with the spectroscopic measurements.
In another case of XTE~J1550$-$564, where a type-C QPO and a high frequency QPO were detected simultaneously in an archival \textit{RXTE} observation, \cite{Motta2014MNRAS.439L..65M} calculated the spin of the BH using RPM and inferring the mass from the OIR spectroscopic measurements. In both sources, the frequencies of the broad noise components were observed to be consistent with the low-frequency extension of the HFQPOs. 
The observations in \citet{Motta2014MNRAS.437.2554M} and \citet{Motta2014MNRAS.439L..65M} are the only observations in the \textit{RXTE} archive in which simultaneous detection of HFQPOs with low frequency QPO was observed, highlighting the rarity of these detections.

\src\ (optical counterpart: ASSASN-18ey) was detected in X-rays on 11$^{\rm{th}}$ March 2018 \citep{2018ATel11399....1K, 2018ATel11403....1K}. The detection of the source by \textit{Gaia} allowed an accurate measurement of the distance of 3.8$^{+2.9}_{-1.2}$~kpc 
\citep{2019MNRAS.485.2642G}.The radio parallax measurement is consistent with optical measurement and places a tighter constraint on the distance of the source \citep[$2.96\pm0.33$~kpc, ][]{Atri2020MNRAS.493L..81A}. It was closely monitored with different space and ground based observatories as it reached high flux levels in LHS and many studies have been published \citep{Shidatsu2018ApJ...868...54S, kara2019Natur.565..198K, Shidatsu2019ApJ...874..183S, Bharali2019MNRAS.487.5946B, Buisson2019MNRAS.490.1350B, Stiele2020ApJ...889..142S, Homan2020ApJ...891L..29H, dzielak2021arXiv210211635D,zdz2021ApJ...909L...9Z, demarco2021arXiv210207811D}.
The emission during the hard states was characterised by a typical accretion disk observed in BHBs and a non-thermal component. The disk temperature of 0.13~keV and the inner disk truncation of 5.1 gravitational radii (\rg = GM/c$^2$, M is the mass of the BH) was observed \citep[][]{Bharali2019MNRAS.487.5946B}.  The non-thermal component (corona)  was modelled with a lamppost geometry \citep{kara2019Natur.565..198K, Bharali2019MNRAS.487.5946B, Buisson2019MNRAS.490.1350B} or a radially distributed corona \citep{zdz2021ApJ...909L...9Z, dzielak2021arXiv210211635D} which yielded different measurements of inner truncation radius ($\sim$5 and  $\sim$10~\rg respectively). 
The soft state analysis of the source using \nus\ is unable to constrain the spin of the source via  reflection spectroscopy \citep{Buisson2021MNRAS.500.3976B}. \citet{guan2020arXiv201212067G} use the continuum modelling of the soft state spectra to estimate the spin of the BH to be $0.2^{+0.2}_{-0.3}$.  

Optical spectroscopy of the source indicates a mass function of 5.18$\pm$0.15 $M_{\odot}$ \citep{Torres2019ApJ...882L..21T} and a mass ratio of 0.072$\pm$0.012 \citep[companion mass divided by compact object mass;][]{Torres2020ApJ...893L..37T}. Using radio observations \citet{Atri2020MNRAS.493L..81A} determine the jet inclination of the source to be 63$\pm$3$^{\circ}$. Using the inclination from radio measurements, \citet{Torres2020ApJ...893L..37T} determine the mass of the BH to be $8.48^{+0.79}_{-0.72}$ $M_{\odot}$. X-ray spectroscopic results also indicate a similar mass but have a wider confidence interval \citep{Shidatsu2018ApJ...868...54S, Bharali2019MNRAS.487.5946B, Chakraborty2020MNRAS.498.5873C}.

\citet{Stiele2020ApJ...889..142S} and \citet{Homan2020ApJ...891L..29H} have reported a comprehensive analysis of the QPOs observed in this particular source. \citet{Stiele2020ApJ...889..142S} describe the QPOs and the variability behaviour throughout the HIMS of the source while \citet{Homan2020ApJ...891L..29H} depict the transition from type-C QPO in HIMS to type-B QPO in SIMS and to the lack of variability in the soft state, which happened in the duration of a single \nicer\ observation.

In this article, we report the measurement of the spin of the BH by applying the RPM to the observed QPOs and broad noise features in the high cadence \nicer\ observations. We describe the observations and methods used to model the PDS in section~\ref{sec:obs} and discuss the results obtained in section~\ref{sec:res}. 

\section{Observations and Data reduction} \label{sec:obs}

\nicer\ \citep{Gendreau2016SPIE.9905E..1HG} extensively monitored the outburst of \src\ over a six months period. The observations were conducted with a cadence of 1--3 days with differing exposures. For each observation, we  combined the data from different MPUs into a single file \texttt{ni1200120*\_ufa.evt} and performed the standard cleaning using \texttt{nicerclean} (\textsc{heasoft version 6.25}). The lightcurves were extracted using \textsc{xselect} while the power density spectra (PDS) were extracted using the  General High-energy Aperiodic Timing Software (\textsc{ghats} version 1.1.0)\footnote{The software can be downloaded from  \url{http://www.brera.inaf.it/utenti/belloni/GHATS/Home.html}}. We extracted the PDS in the energy range of 0.01-12~keV\footnote{The 0.01-0.2 energy band has a prominent noise peak in the energy spectrum but has negligible contribution in the PDS which is why the energy band is typically ignored in the spectral analysis but can be used for timing studies}, using a minimum time resolution of 0.0004~s (probing the Nyquist frequency of 1250~Hz). The PDS were created from continuous light curves of 26.2144~s (corresponds to $2^{16}$ bins), which are averaged for each observation. The averaged PDS was rebinned logarithmically such that each bin is $e^{0.01}$ times the previous bin in duration. Some of the 26.2144~s segments in different observations had spuriously low count rates due to data drop and the PDS from these segments were excluded from the averaging. 

The higher frequencies ($\gtrsim$20~Hz) were dominated by the Poisson noise and due to the deadtime effects, the power  was observed to be  slightly less than 2 in the Leahy normalised PDS \citep{Leahy1983ApJ...266..160L}. 
The reduction in power due to deadtime is flux dependent and thus has to be computed accordingly. In the present study, we model the effects of Poissonian noise in the PDS by including a zero-slope power law component in the PDS modelling. 

The PDS of hard state observations are distinctly different from the soft state PDS, which is typical of BHBs \citep{belloni1999ApJ...519L.159B}. Since our focus is on the evolution of the PDS, we only consider the observations which show significant power (total fractional rms $\gtrsim$10\%) after Poisson noise subtraction in the frequency range of 0.03--1250~Hz. The observations analysed in the work are tabulated in Table~\ref{tab:obs}. 

\subsection{PDS modelling}

\begin{figure}
    \centering
    \includegraphics[width=0.45\textwidth]{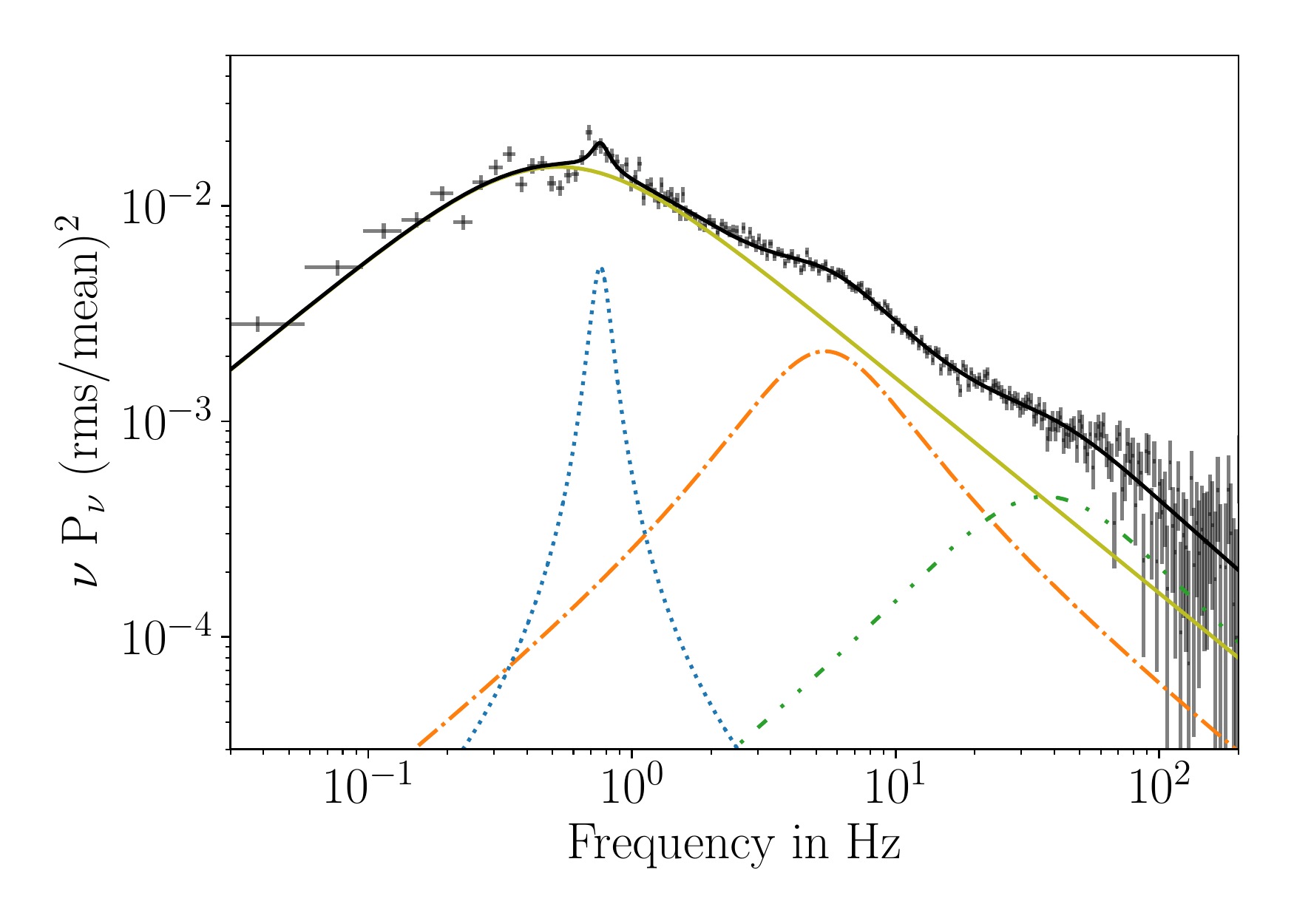}
    \caption{Power density spectrum of \src\ as observed by \nicer\ for the observation ID 1200120161. The PDS is normalised to fractional rms squared units after subtraction of Poissonian noise. For plotting purpose, we have multiplied the power with the frequency to highlight the position of the characteristic frequencies of different components. The PDS was modelled using Lorentzian function (individual components are shown in different colours). The detected QPO is shown as blue dotted line while the associated broad noise components are shown as yellow solid, orange dash-dotted and green dash-double dotted lines respectively.}
    \label{fig:pds}
\end{figure}

The hard-state PDS were converted to \textsc{xspec} readable format using \textsc{ghats}. The PDS were phenomenologically modelled using Lorentzian components \citep[][]{Belloni2002ApJ...572..392B}. For the features with width (\textit{w}) much greater than the centroid frequency ($\nu_{c}$), $\nu_{c}$ was frozen at 0 and these are referred to as Broad Low Noise (BLN) features.
A PDS for one of the observations is shown in Figure~\ref{fig:pds}. The components required to model the PDS are shown in different colours. The Poissonian noise has been subtracted from the PDS and the PDS have been renormalised to fractional rms squared units \citep{belloni1990A&A...227L..33B}. 

\begin{figure}
    \centering
    \includegraphics[width=0.45\textwidth]{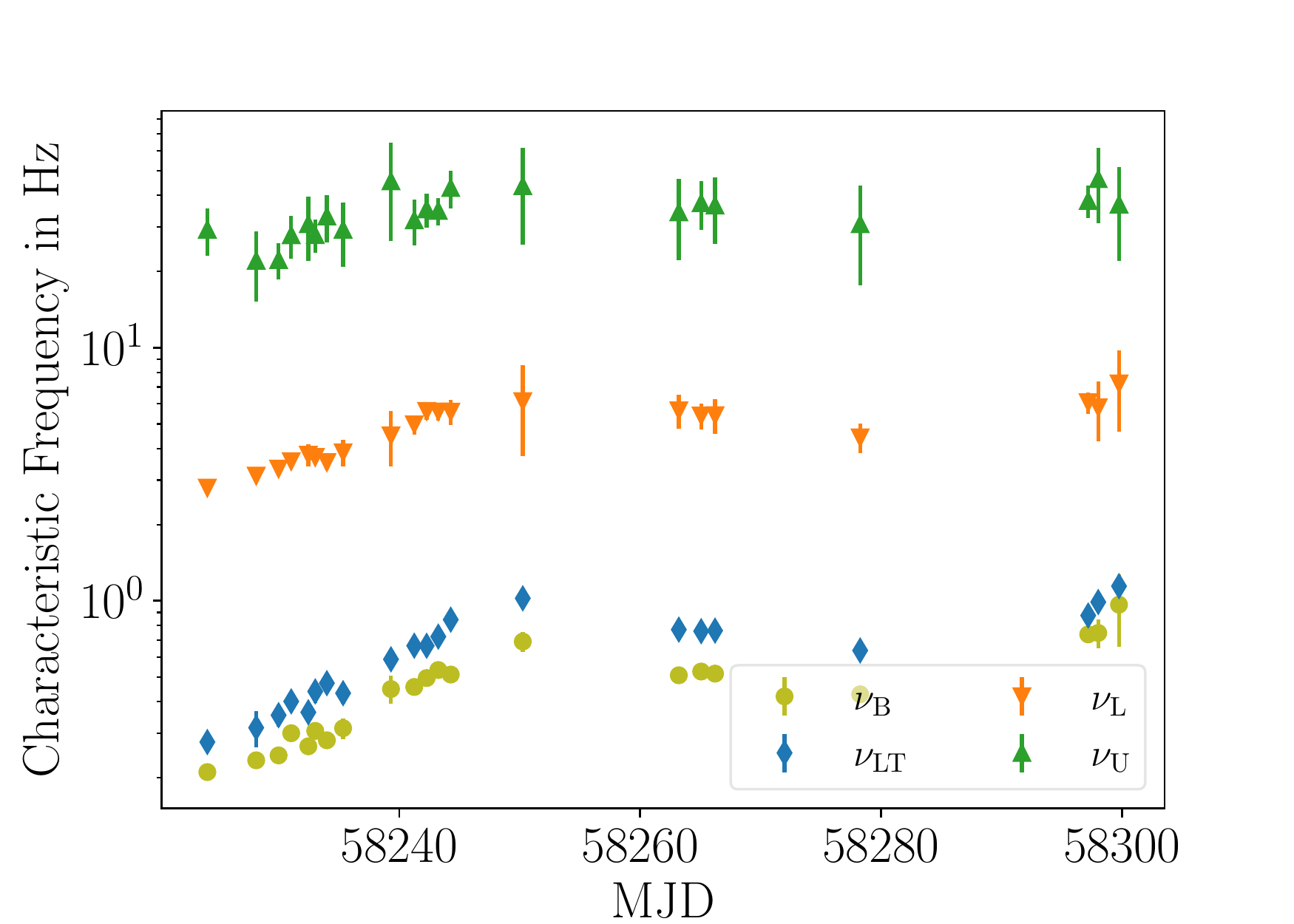}
    \caption{The evolution of the characteristic frequencies of different components is shown here. The colour scheme of the components is kept identical to Figure~\ref{fig:pds}. The $1\sigma$ error bars are also indicated for all the frequencies. }
    \label{fig:char_freq_vs_mjd}
\end{figure}

\begin{figure}
    \centering
    \includegraphics[width=0.47\textwidth]{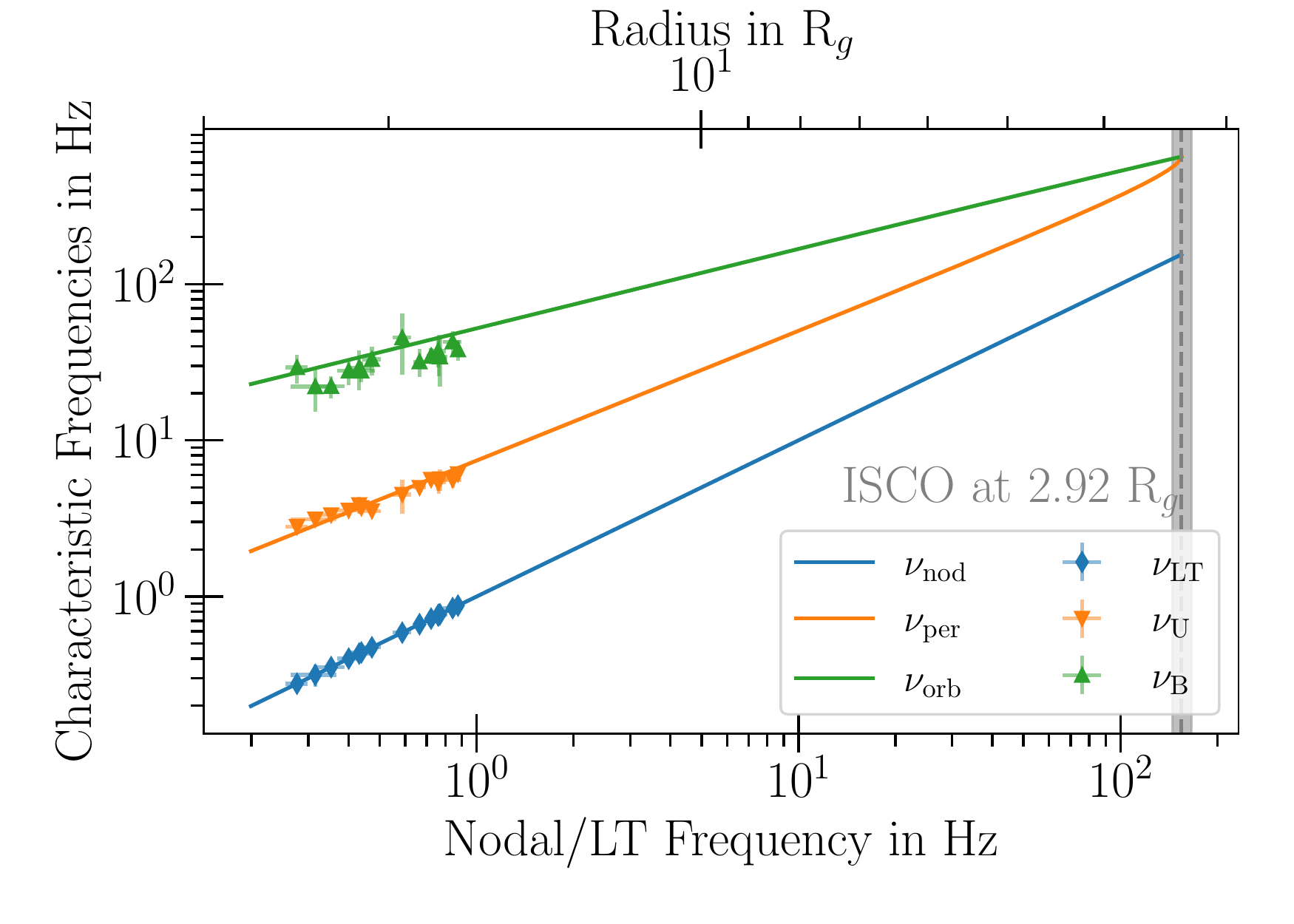}
    \caption{The correlation between the characteristic frequencies of different components with $\nu_{\rm{LT}}$ is shown here. The colour scheme of the points is kept consistent with features observed in Figures~\ref{fig:pds} and \ref{fig:char_freq_vs_mjd}. The correlation of the blue points (i.e. $\nu_{\rm{LT}}$) is artificial as the same points are plotted with each other. The solid lines correspond to the frequencies predicted by the best fit RPM. The vertical dashed line corresponds to the highest QPO frequency that can be expected  for the assumed mass and estimate spin of the source. The grey band corresponds to the 1$\sigma$ confidence interval on the highest QPO frequency. 
    }
    \label{fig:corr_rpm}
\end{figure}

A subset of the hard state observations show a prominent low frequency QPO. The detections of the QPO are also reported in \citet{Stiele2020ApJ...889..142S} and \citet{Mudambi2020ApJ...889L..17M}. \cite{Stiele2020ApJ...889..142S} have detected and reported a pair of QPOs throughout the \nicer\ observations with varying quality factors and significance. The pair of the QPOs are harmonically linked harmonically linked and the higher frequency QPO typically is more significant. In our analysis, we model only the higher frequency QPO as it is stronger and the addition of the lower frequency QPO did not change the fit statistic significantly. The observations which show the QPO have been tabulated in Table \ref{tab:obs} along with the observed QPOs and the characteristic frequencies of the higher broad noise components. To compare with the previous works \citep[e.g.][]{Belloni2002ApJ...572..392B, Motta2014MNRAS.437.2554M, Motta2014MNRAS.439L..65M}, we label the detected QPOs as $\nu_{\rm{LT}}$ and the lowest, middle and higher broad noise components as $\nu_{\rm{B}}$, $\nu_{\rm{L}}$ and $\nu_{\rm{U}}$ respectively. The variation of the QPOs and other components is shown in  Figure~\ref{fig:char_freq_vs_mjd}.  The QPOs detected by \citet{Stiele2020ApJ...889..142S} and this work match within the 1$\sigma$ confidence interval. 
These observations also show broad noise components similar to the ones observed in \citet{PBK1999ApJ...520..262P}, \citet{Belloni2002ApJ...572..392B}, \citet{Motta2014MNRAS.437.2554M} and \citet{Motta2014MNRAS.439L..65M}. The frequencies at which these broad features peak in the $\nu\rm{P}_\nu$ plot \citep[also known as the characteristic frequency; $\nu_{\rm{char}} = \sqrt{\nu_{\rm{c}}^2+(w/2)^2}$, ][]{Belloni2002ApJ...572..392B} and the QPO frequency are correlated with each other (see figure~\ref{fig:char_freq_vs_mjd}). 
We also plot $\nu_{\rm{char}}$ of different components with the QPO frequency in figure \ref{fig:corr_rpm} to highlight the correlation of all the frequencies.

\begin{table*}
\centering
    \caption{Summary of the QPOs and broad noise features used to constrain the spin of the source. The error bars reported here correspond to 1$\sigma$ confidence interval. }
    \label{tab:obs}
    \begin{tabular}{c|cccc}
    \hline
Observation ID & Start MJD &  $\nu_{\rm{LT}}$ (Hz) &  $\nu_{\rm{L}}$ (Hz) & $\nu_{\rm{U}}$ (Hz) \\ \hline
1200120130 & 58224.068 & $0.276 \pm 0.021$ & $2.79 \pm 0.09$ & $29.2 \pm  6.1$ \\
1200120132 & 58228.124 & $0.315 \pm 0.051$ & $3.12 \pm 0.18$ & $22.0 \pm  6.8$ \\
1200120134 & 58229.984 & $0.353 \pm 0.035$ & $3.32 \pm 0.12$ & $22.2 \pm  3.6$ \\
1200120135 & 58231.032 & $0.400 \pm 0.031$ & $3.56 \pm 0.20$ & $27.8 \pm  5.3$ \\
1200120136 & 58232.452 & $0.362 \pm 0.014$ & $3.77 \pm 0.37$ & $30.7 \pm  8.7$ \\
1200120137 & 58233.026 & $0.438 \pm 0.043$ & $3.69 \pm 0.18$ & $27.8 \pm  4.2$ \\
1200120138 & 58233.991 & $0.473 \pm 0.031$ & $3.52 \pm 0.23$ & $33.0 \pm  6.9$ \\
1200120139 & 58235.339 & $0.431 \pm 0.039$ & $3.86 \pm 0.47$ & $29.2 \pm  8.3$ \\
1200120141 & 58239.314 & $0.587 \pm 0.038$ & $4.50 \pm 1.10$ & $45.4 \pm 19$ \\
1200120143 & 58241.246 & $0.664 \pm 0.029$ & $4.97 \pm 0.43$ & $31.8 \pm  6.5$ \\
1200120144 & 58242.275 & $0.664 \pm 0.016$ & $5.61 \pm 0.43$ & $35.2 \pm  5.4$ \\
1200120145 & 58243.242 & $0.722 \pm 0.024$ & $5.60 \pm 0.45$ & $34.7 \pm  4.2$ \\
1200120146 & 58244.276 & $0.842 \pm 0.055$ & $5.58 \pm 0.63$ & $42.7 \pm  7.4$ \\
1200120152 & 58250.255 & $1.023 \pm 0.070$ & $6.14 \pm 2.41$ & $43.5 \pm 18$ \\
1200120159 & 58263.209 & $0.769 \pm 0.040$ & $5.65 \pm 0.88$ & $34.2 \pm 12$ \\
1200120161 & 58265.064 & $0.758 \pm 0.042$ & $5.39 \pm 0.62$ & $37.2 \pm  8.1$ \\
1200120162 & 58266.217 & $0.762 \pm 0.024$ & $5.41 \pm 0.84$ & $36.4 \pm 11$ \\
1200120174 & 58278.263 & $0.636 \pm 0.025$ & $4.41 \pm 0.58$ & $30.7 \pm 13$ \\
1200120189 & 58297.194 & $0.875 \pm 0.013$ & $6.08 \pm 0.59$ & $38.0 \pm  5.6$ \\
1200120190 & 58298.028 & $0.989 \pm 0.028$ & $5.81 \pm 1.55$ & $46.3 \pm 15$ \\
1200120191 & 58299.752 & $1.143 \pm 0.062$ & $7.22 \pm 2.55$ & $36.8 \pm 15$ \\
         \hline
    \end{tabular}
    
\end{table*}

\begin{figure}
    \centering
    \includegraphics[width=0.49\textwidth]{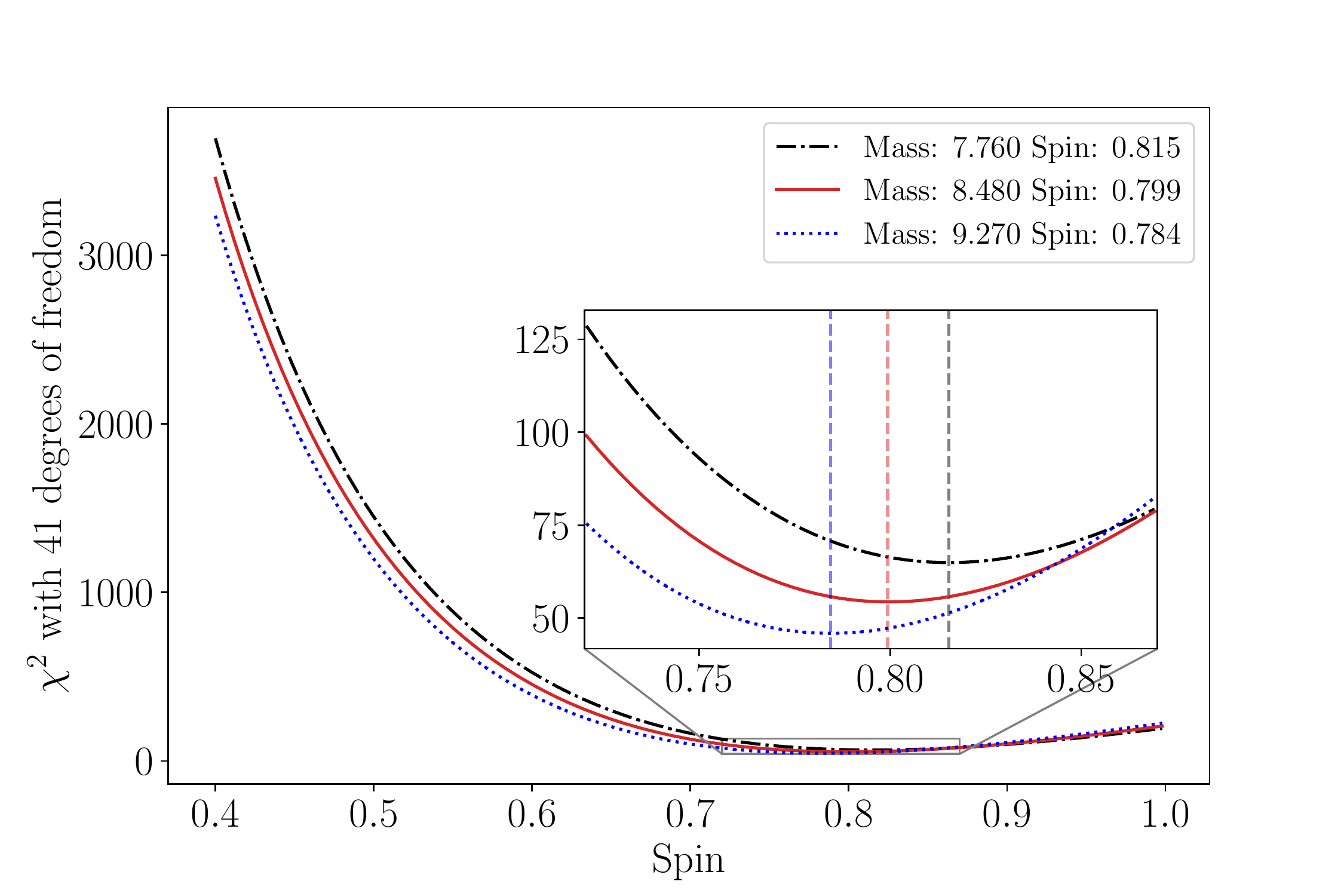}
    \caption{Variation of total $\chi^2$ with the spin for different assumed masses of the source. The choice of masses is from the current mass estimate of the source from \citet{Torres2020ApJ...893L..37T} and corresponds to the 1$\sigma$ lower limit (shown in black dot dashed line), best guess (shown in red solid line) and 1$\sigma$ upper limit (shown in blue dotted line). The inset shows variation near the minima of the $\chi^2$ and vertical dashed lines indicate the minimum of each assumed mass. The predicted spin values for each mass is reported in the legend.}
    \label{fig:chi2_plot}
\end{figure}

As seen in \citet{Motta2014MNRAS.437.2554M} and \citet{Motta2014MNRAS.439L..65M}, the characteristic frequencies of the broad features follow the trend predicted by RPM, although  with a significant scatter. Such a scatter is likely related to the fact that the physics of the accretion flow is more complex than that assumed by the RPM, as well as to a more practical reason, namely that the characteristic frequency of a broad PDS component is not an obvious measurable, and the definition we assumed might not be accurate enough. The implication of the above will be discussed in Sec. \ref{sec:res}. In principle, one could use triplets formed by two broad PDS components (associated with the high-frequency QPOs, see \citealt{PBK1999ApJ...520..262P} and \citealt{Motta2014MNRAS.437.2554M} and a low-frequency QPO to constrain the mass and spin of a BH. However, given the presence of a large scatter, using triplets individually could yield inconsistent values of the mass and spin. In order to mitigate the effects of the scatter,  we can fit the trend followed by the frequencies to determine the optimal parameters of the BH, which will be therefore estimated based on the overall correlation, rather than on individual (possibly biased) points. 
Using optical observations and jet inclination, \citet{Torres2020ApJ...893L..37T} have determined the mass of the source which we have used as an input. Using the equation of nodal precession frequency from \citet{Motta2014MNRAS.437.2554M}, we solve for the radius of oscillation for an assumed spin of the black hole using the Newton-Raphson method.  We compute the frequency of the periastron precession and orbital oscillation frequency using the equations from \citet{Motta2014MNRAS.437.2554M} at that radius. We compute the $\chi^2$ between the model frequencies and the characteristic frequencies of the observed broad noise components. Since individual detections of frequencies are independent, we summed the $\chi^2$ from each observation to obtain a total $\chi^2$. Varying the spin as a parameter, we determine the spin of the source where the total $\chi^2$ is found to be minimum. The variation of the total $\chi^2$ with the spin is shown in Figure~\ref{fig:chi2_plot}. We repeated the process for different mass values indicated by the 1$\sigma$ confidence interval from \citet{Torres2020ApJ...893L..37T}. The correlation in  Figure~\ref{fig:corr_rpm} is also overplotted with the predicted RPM frequencies for the assumed mass and the computed spin. The radius of oscillation corresponding to the QPO is shown in the top axis of  Figure~\ref{fig:corr_rpm}.

\begin{figure}
    \centering
    \includegraphics[width=0.8\columnwidth]{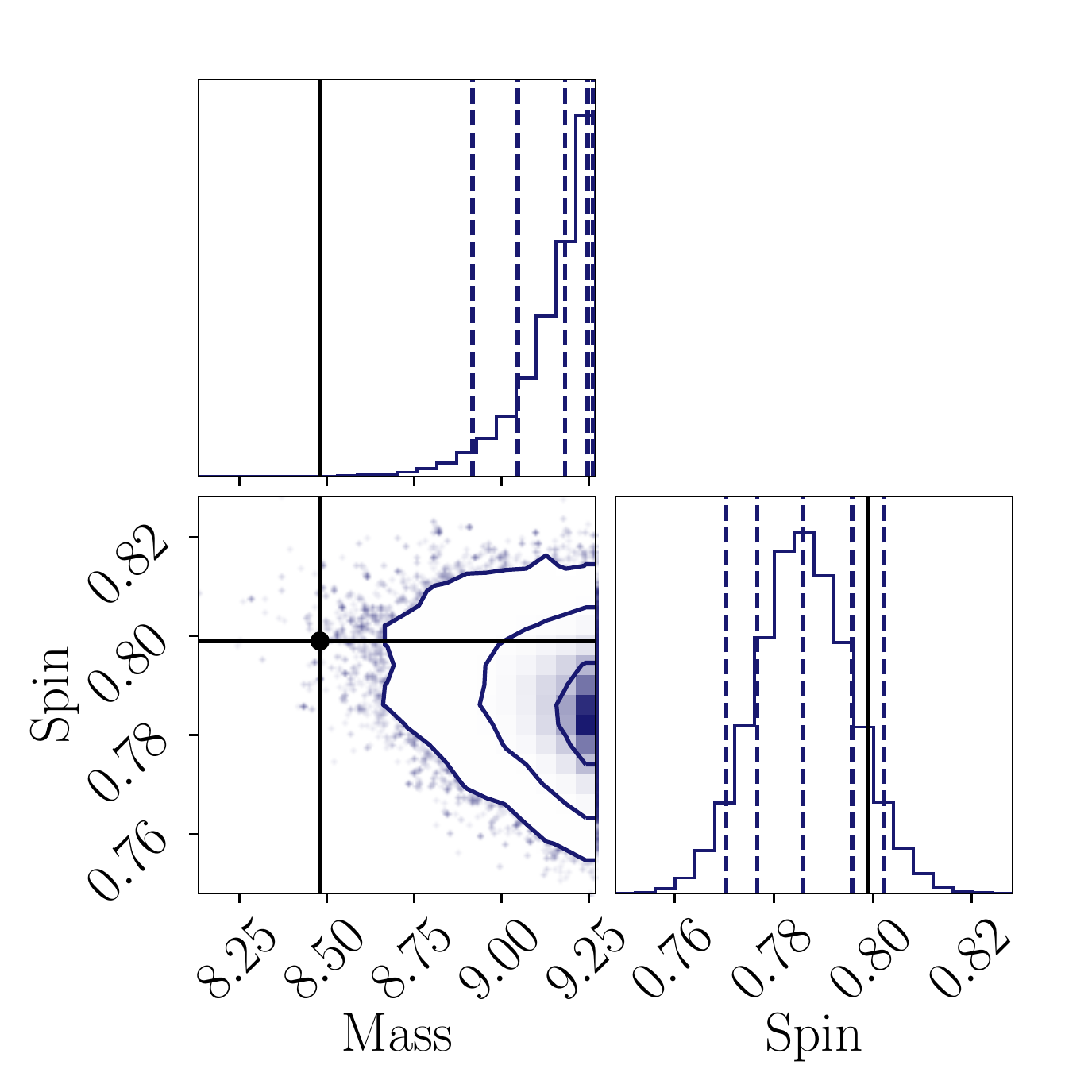}
    \includegraphics[width=0.8\columnwidth]{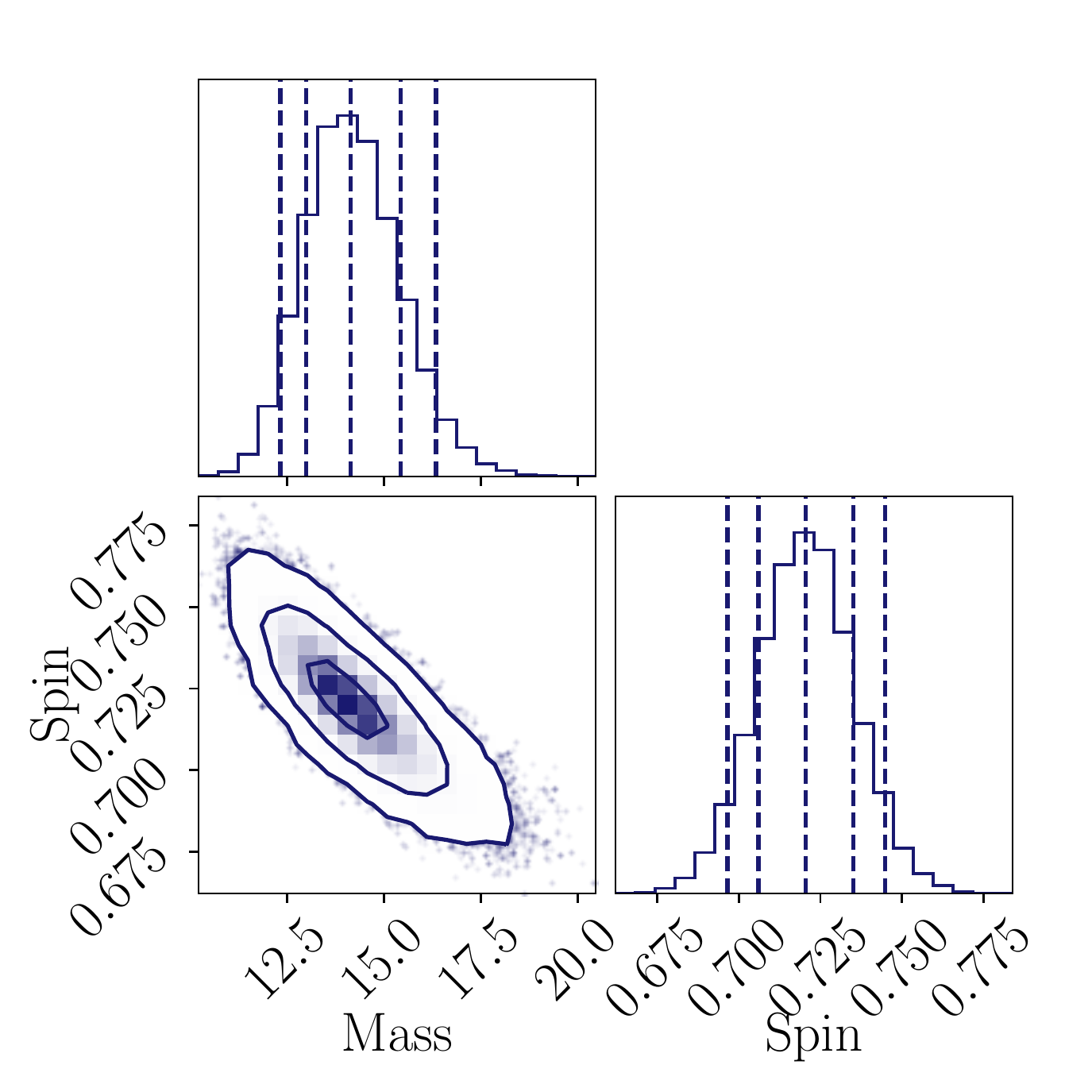}
    \includegraphics[width=0.8\columnwidth]{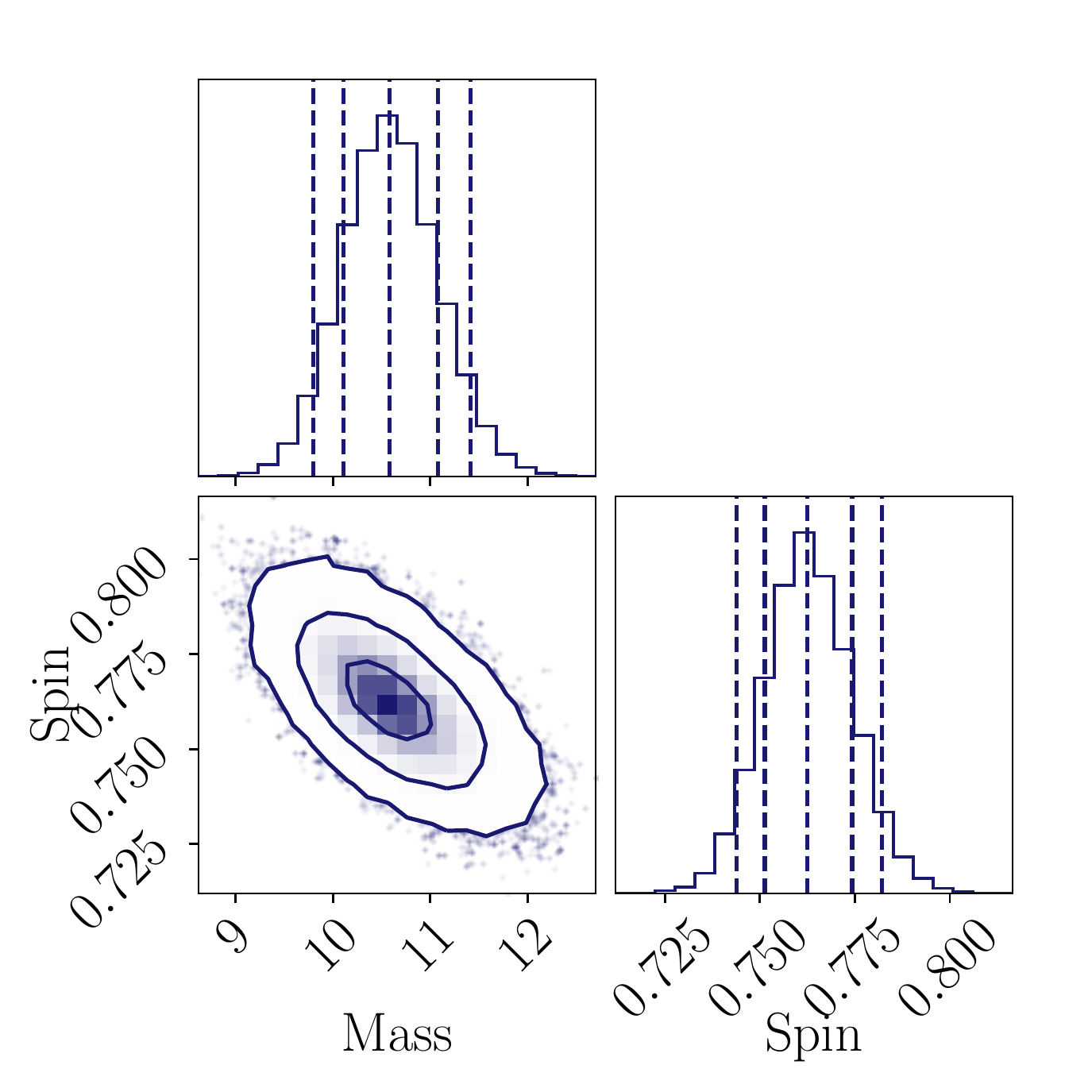}
    \caption{Corner plots for MCMC sampling of parameter space. The top panel considers a uniform prior over 1$\sigma$ confidence interval from \citet{Torres2020ApJ...893L..37T}. The black point indicates the spin estimate from the $\chi^2$ minimisation for assumed mass. The middle plot considers a uniform prior over a wide mass range of 5--20 M$_\odot$ while the bottom plot assumes a Gaussian prior on mass with 8.48~M$_\odot$ as mean and 0.7~M$_\odot$ as the standard deviation. In all three cases the spin is drawn from a uniform distribution ranging from -0.998 to 0.998.  }
    \label{fig:mass_spin}
\end{figure}

We find that the $\chi^2$ is minimum (54.354 for 41 degrees of freedom) for the spin value of 0.799 when the assumed mass is 8.48~M$_\odot$ \citep[central value in the interval of][]{Torres2020ApJ...893L..37T}. The lowest $\chi^2$ for the lower and upper limit of mass interval from \citet{Torres2020ApJ...893L..37T} corresponds to a spin of 0.815 and 0.784 respectively. The uncertainty on the spin measurement from the $\chi^2$ distribution is lower that the one propagated from the mass confidence interval by an order of magnitude and thus we report the uncertainty in spin as determined from the error propagation from the mass confidence interval.  
We also determine the variation of the $\chi^2$ as a function of both mass and spin.
Assuming the likelihood as determined from the $\chi^2$ method, we apply a uniform prior on mass and spin. For spin we allow for complete range i.e. -0.998 to 0.998 and for mass we first assume the uniform prior spanning  1$\sigma$ interval from \citet{Torres2020ApJ...893L..37T} and then we assume  an extremely conservative uniform prior of 5--20~M$_\odot$. We used Markov-Chain Monte Carlo sampling using \textsw{emcee} package \citep[][]{emcee}   by initialising 150 walkers around a narrow interval around the guess mass and spin of 8.48~M$_\odot$ and 0.8 respectively. The walkers were allowed to move 1500 steps individually and initial 200 steps were discarded to remove the effects of initialisation. As a second test, we apply a Gaussian prior on the mass assuming a mean of 8.48 M$_\odot$ and a standard deviation of 0.7 M$_\odot$ \citep[as suggested by ][]{Torres2020ApJ...893L..37T}
We have plotted the 2-D probability distribution of  mass and spin and the marginalised distributions of mass and spin for both cases in Figure \ref{fig:mass_spin}. The spin measurement for a fixed mass of 8.48~M$_\odot$ is indicated as a black point in the top panel.

\section{Results and Discussion} \label{sec:res}

We used observations of the BHB \src\ with \nicer\ to study the variation of the timing properties of the source. The PDS of the source showed significant broadband noise in 0.03--100~Hz. Some of the observations also show a type-C QPO (typically $<$1~Hz). These observations also show broadband noise features at higher frequencies which correlate with the evolution of the QPO as seen in other X-ray binaries \citep{PBK1999ApJ...520..262P, Belloni2002ApJ...572..392B, Motta2014MNRAS.437.2554M, Motta2014MNRAS.439L..65M}. 

We fit the observed QPO and the characteristic frequencies of the broad noise features to the Relativistic Precession model. We use the mass of the BH inferred from the spectro-photometric optical observations to put tight constraints on the spin of the source. We derive a spin of the source of $0.799^{+0.016}_{-0.015}$.  The ISCO corresponding to this spin is 2.91~\rg. The low frequency QPOs corresponds to large radii of oscillation (15--25~\rg, see Figure \ref{fig:corr_rpm}). Allowing the mass and spin to vary simultaneously, the mass of the source is overestimated, as compared to \citet{Torres2020ApJ...893L..37T} measurement, by a factor of 1.2 or 1.6 (depending on the choice of prior on the mass, Gaussian or Uniform). For the overestimated mass, the corresponding spin estimate is reduced to 0.76 or 0.72 respectively. Restricting the mass to 1$\sigma$ confidence interval suggested by \citet{Torres2020ApJ...893L..37T} prefers a 9.25~M$_\odot$ BH with a spin of 0.78. 

\subsection{RPM and broad components}

\citet{PBK1999ApJ...520..262P} have associated the broad noise components as the low frequency counterparts of the HFQPOs. \citet{Motta2014MNRAS.437.2554M} and \citet{Motta2014MNRAS.439L..65M} have observed a similar trend for GRO~J1655$-$40 and XTE~J1550$-$564 respectively. The broad components are observed simultaneously with a low frequency type-C QPO ($\leq$1~Hz), which occur at a larger radius. In XTE~J1550$-$564, some of the \textit{RXTE} observations show the presence of broad components which match the predicted periastron precession and orbital frequencies motivating the hypothesis that the broad features are indeed the low frequency counterparts. The broad components in these sources have shown a scatter around the trend predicted by RPM. In \nicer\ observations of \src, we see a similar phenomenon, which has motivated us to use RPM to describe the observed frequencies. \citet{Motta2014MNRAS.439L..65M} discusses the possible reasons for the scatter of the broad noise components. The broad components are associated with the oscillations arising at a larger radii and at these radii the test particle orbits might differ substantially from the accretion orbits. The scatter between the components can also arise from the assumptions in the modelling of these features.

While the mass posterior we recover is consistent with the mass value obtained by \cite{Torres2020ApJ...893L..37T}, the central mass is significantly larger than the best value reported by such authors. This effect might be due to a bias introduced by the fact that in this work we consider broad PDS components rather than QPOs, which do not provide a clear centroid frequency. 
According to the PBK relation \citep{PBK1999ApJ...520..262P}, certain broad components in the PDS are associated to HFQPOs when these are generated at large distances from the central BH.  By fitting large PDS components with the RPM, rather than narrow HFQPOs (which are instead assumed to be generated at small radii from the BH) results are much more affected by the effects of matter precessing as part of an extended disc \citep{Motta2018MNRAS.473..431M}. On the one hand, adopting the RPM at large radii, one models an extended disc assuming it is a narrow precessing ring. This implies that the radii are assumed to be smaller than they really are (for a given radius, the extended disc precession frequency is larger than a ring precession frequency). This induces a shift to higher masses (and smaller spins), which counterbalances the bias in radius. On the other hand, as one considers wider and wider orbits, the torque exerted by the outer disc on a particle on a given orbit grows larger compared to the torque exerted by the frame dragging, so that precession might be slowed down. This means that what is actually measured is a precession frequency slower than expected, which once more induces a shift towards larger masses. These two effects combined might effectively push the posterior distribution to higher masses.

\subsection{Spin measurement}

The ISCO corresponding to spin of 0.8 is 2.91~\rg, which allows for a truncated inner disk at $\sim$5~\rg\ \citep{Bharali2019MNRAS.487.5946B, Buisson2019MNRAS.490.1350B}. In their analysis, \citet{Bharali2019MNRAS.487.5946B} found that the broadband spectra is consistent with a maximal spinning BH with a lower limit of 0.68, which is consistent with our measurements.  The observation of inner truncated disk at the 5.3~\rg\ is interpreted as the ISCO by \citet{Buisson2019MNRAS.490.1350B} which corresponds to a low spin but in the hard state of BHBs, the accretion disk is typically truncated and need not correspond to ISCO \citep{Done2007A&ARv..15....1D}. 
\cite{Fabian2020MNRAS.493.5389F} have reported a strong degeneracy between the spin and inclination measurement. A high inclination of the source  \citep[from radio measurements][]{Atri2020MNRAS.493L..81A} would correspond to a retrograde spin of the black hole while a lower inclination \citep[$\sim30^\circ$][]{Bharali2019MNRAS.487.5946B, Buisson2019MNRAS.490.1350B} of the source supports a high prograde spin. A retrograde spin is quite inconsistent with the measured inner truncation radius of $\sim5$~\rg. On the other hand, our spin measurement would imply that the inclination of the disk is closer to $30^\circ$ and indicate a possible misalignment between the inner disk inclination and the jet inclination \citep{MJ2019Natur.569..374M}. The spectral continuum modelling in the soft state of the source by \citet{guan2020arXiv201212067G} yields a low spin of the source which we note significantly differs from our measurement. The continuum modelling method assumes that the inner truncation is  reaching the ISCO which is unconfirmed. The measured inner truncation can thus only provide a lower limit on the spin of the source and is still consistent with a high spin measurement.   

As indicated in the previous section, the method overestimates the mass of the BH due to the interpretation of the broad features arising from a narrow ring. The overestimation of the mass implies an underestimation of the spin as indicated by the contours in figure \ref{fig:mass_spin}. Even with fixing the mass at 8.48 M$_\odot$, the estimated spin of 0.799 may be regarded as a lower limit owing to the assumption that the broad noise features behave strictly according to the RPM. The  quantification of the bias inherent in this method will be followed up in a future work.

\subsection{Disk truncation}

The measurement of the inner disk truncation from the spectroscopic measurements vary significantly as different accretion models have different inherent geometrical and physical assumptions leading to the discrepancy. Most of the works have analysed the hard or hard-intermediate state observations making it easier to compare and track the evolution.
\citet{dzielak2021arXiv210211635D} utilise the hard state observation and place the inner truncation radius at 45~\rg\ using the temperature constraints from the frequency resolved spectrum. In contrast \citet{Bharali2019MNRAS.487.5946B} determine the inner truncation radius of the source at 5.1~\rg\ at a similar epoch. \citet{Buisson2019MNRAS.490.1350B} have analysed the hard and hard-intermediate state observations and have observed a truncation at 5.3~\rg, while \citet{zdz2021ApJ...909L...9Z} and \citet{demarco2021arXiv210207811D} suggests that in the hard and hard intermediate states, the inner truncation of the disk is $\gtrsim$10~\rg. 
The observations we have analysed in the present work correspond to hard intermediate states stopping a few days prior to the state transition discussed in \citet{Homan2020ApJ...891L..29H} and \citet{demarco2021arXiv210207811D}. The \textit{special} radius at which QPO is arising, as suggested by the RPM, varies within 15--25~\rg. This radius although inconsistent with the truncation radius from spectral measurements assuming a lamppost geometry \citep{Bharali2019MNRAS.487.5946B, Buisson2019MNRAS.490.1350B}, is allowed by alternative geometries \citep{demarco2021arXiv210207811D, zdz2021ApJ...909L...9Z, dzielak2021arXiv210211635D}. 
In these geometries, the special radius at which QPO is arising could correspond to a transition region to a hot accretion plasma \citep{dzielak2021arXiv210211635D} or could indicate the extent of covering by the comptonising medium \citep{zdz2021ApJ...909L...9Z}.

\section{Conclusions}

Using the RPM, we provide an estimate of the spin of the source at 0.799 $^{+0.016}_{-0.015}$ for an assumed mass of the source at 8.48~M$_\odot$. Exploring a wider parameter space indicates that the method is sensitive to assumption of the priors on the mass and for uninformative priors, the method prefers a mass higher than the one estimated by optical measurements and a slightly lower spin. However, we argue that the use of broad features instead of QPO peaks in the analysis of the Relativistic Precession Model tends to introduce a bias so as to underestimate the spin and overestimate the mass.  The estimated spin quoted above should therefore be regarded as a lower limit.

\section*{Data Availability}

The data underlying this article are available in \url{https://heasarc.gsfc.nasa.gov/docs/archive.html} using the observation IDs listed in Table \ref{tab:obs}. The associated software to process the data are also available publicly at the same. TMB can be contacted for \textsc{GHATS}. 

\section*{Acknowledgements}
This work was supported by NASA through the \nicer\ mission and the Astrophysics Explorers Program. This research has made use of data, software, and/or web tools obtained from the High Energy Astrophysics Science Archive Research Center (HEASARC), a service of the Astrophysics Science Division at NASA/GSFC and of the Smithsonian Astrophysical Observatory's High Energy Astrophysics Division. 
YB would like to acknowledge the support received from CSIR. This work has been supported
by the Executive Programme for Scientific and Technological cooperation between the Italian Republic and the Republic of India for the years 2017-2019 under project IN17MO11 [INT/Italy/P11/2016 (ER)]. TMB acknowledges support from the agreement ASI-INAF n.2017-14-H.0 and PRIN-INAF 2019 n.15. GP acknowledges funding from the European Research Council (ERC) under the European Union's Horizon 2020 research and innovation programme (grant agreement No. 865637). 




\bibliographystyle{mnras}
\bibliography{ref.bib} 

\begin{thebibliography}{}
\makeatletter
\relax
\def\mn@urlcharsother{\let\do\@makeother \do\$\do\&\do\#\do\^\do\_\do\%\do\~}
\def\mn@doi{\begingroup\mn@urlcharsother \@ifnextchar [ {\mn@doi@}
  {\mn@doi@[]}}
\def\mn@doi@[#1]#2{\def\@tempa{#1}\ifx\@tempa\@empty \href
  {http://dx.doi.org/#2} {doi:#2}\else \href {http://dx.doi.org/#2} {#1}\fi
  \endgroup}
\def\mn@eprint#1#2{\mn@eprint@#1:#2::\@nil}
\def\mn@eprint@arXiv#1{\href {http://arxiv.org/abs/#1} {{\tt arXiv:#1}}}
\def\mn@eprint@dblp#1{\href {http://dblp.uni-trier.de/rec/bibtex/#1.xml}
  {dblp:#1}}
\def\mn@eprint@#1:#2:#3:#4\@nil{\def\@tempa {#1}\def\@tempb {#2}\def\@tempc
  {#3}\ifx \@tempc \@empty \let \@tempc \@tempb \let \@tempb \@tempa \fi \ifx
  \@tempb \@empty \def\@tempb {arXiv}\fi \@ifundefined
  {mn@eprint@\@tempb}{\@tempb:\@tempc}{\expandafter \expandafter \csname
  mn@eprint@\@tempb\endcsname \expandafter{\@tempc}}}

\bibitem[\protect\citeauthoryear{{Abramowicz} \& {Klu{\'z}niak}}{{Abramowicz}
  \& {Klu{\'z}niak}}{2001}]{abramowicz2001A&A...374L..19A}
{Abramowicz} M.~A.,  {Klu{\'z}niak} W.,  2001, \mn@doi [\aap]
  {10.1051/0004-6361:20010791}, \href
  {https://ui.adsabs.harvard.edu/abs/2001A&A...374L..19A} {374, L19}

\bibitem[\protect\citeauthoryear{{Atri} et~al.,}{{Atri}
  et~al.}{2020}]{Atri2020MNRAS.493L..81A}
{Atri} P.,  et~al., 2020, \mn@doi [\mnras] {10.1093/mnrasl/slaa010}, \href
  {https://ui.adsabs.harvard.edu/abs/2020MNRAS.493L..81A} {493, L81}

\bibitem[\protect\citeauthoryear{{Belloni} \& {Hasinger}}{{Belloni} \&
  {Hasinger}}{1990}]{belloni1990A&A...227L..33B}
{Belloni} T.,  {Hasinger} G.,  1990, \aap, \href
  {https://ui.adsabs.harvard.edu/abs/1990A&A...227L..33B} {227, L33}

\bibitem[\protect\citeauthoryear{{Belloni} \& {Motta}}{{Belloni} \&
  {Motta}}{2016}]{belloni2016ASSL..440...61B}
{Belloni} T.~M.,  {Motta} S.~E.,  2016, {Transient Black Hole Binaries}.
p.~61, \mn@doi{10.1007/978-3-662-52859-4_2}

\bibitem[\protect\citeauthoryear{{Belloni}, {M{\'e}ndez}, {van der Klis},
  {Lewin}  \& {Dieters}}{{Belloni} et~al.}{1999}]{belloni1999ApJ...519L.159B}
{Belloni} T.,  {M{\'e}ndez} M.,  {van der Klis} M.,  {Lewin} W.~H.~G.,
  {Dieters} S.,  1999, \mn@doi [\apjl] {10.1086/312130}, \href
  {https://ui.adsabs.harvard.edu/abs/1999ApJ...519L.159B} {519, L159}

\bibitem[\protect\citeauthoryear{{Belloni}, {Psaltis}  \& {van der
  Klis}}{{Belloni} et~al.}{2002}]{Belloni2002ApJ...572..392B}
{Belloni} T.,  {Psaltis} D.,   {van der Klis} M.,  2002, \mn@doi [\apj]
  {10.1086/340290}, \href
  {https://ui.adsabs.harvard.edu/abs/2002ApJ...572..392B} {572, 392}

\bibitem[\protect\citeauthoryear{{Belloni}, {Homan}, {Casella}, {van der Klis},
  {Nespoli}, {Lewin}, {Miller}  \& {M{\'e}ndez}}{{Belloni}
  et~al.}{2005}]{belloni2005A&A...440..207B}
{Belloni} T.,  {Homan} J.,  {Casella} P.,  {van der Klis} M.,  {Nespoli} E.,
  {Lewin} W.~H.~G.,  {Miller} J.~M.,   {M{\'e}ndez} M.,  2005, \mn@doi [\aap]
  {10.1051/0004-6361:20042457}, \href
  {https://ui.adsabs.harvard.edu/abs/2005A&A...440..207B} {440, 207}

\bibitem[\protect\citeauthoryear{{Bharali}, {Chauhan}  \& {Boruah}}{{Bharali}
  et~al.}{2019}]{Bharali2019MNRAS.487.5946B}
{Bharali} P.,  {Chauhan} J.,   {Boruah} K.,  2019, \mn@doi [\mnras]
  {10.1093/mnras/stz1686}, \href
  {https://ui.adsabs.harvard.edu/abs/2019MNRAS.487.5946B} {487, 5946}

\bibitem[\protect\citeauthoryear{{Buisson} et~al.,}{{Buisson}
  et~al.}{2019}]{Buisson2019MNRAS.490.1350B}
{Buisson} D.~J.~K.,  et~al., 2019, \mn@doi [\mnras] {10.1093/mnras/stz2681},
  \href {https://ui.adsabs.harvard.edu/abs/2019MNRAS.490.1350B} {490, 1350}

\bibitem[\protect\citeauthoryear{{Buisson} et~al.,}{{Buisson}
  et~al.}{2021}]{Buisson2021MNRAS.500.3976B}
{Buisson} D.~J.~K.,  et~al., 2021, \mn@doi [\mnras] {10.1093/mnras/staa3510},
  \href {https://ui.adsabs.harvard.edu/abs/2021MNRAS.500.3976B} {500, 3976}

\bibitem[\protect\citeauthoryear{{Casella}, {Belloni}  \& {Stella}}{{Casella}
  et~al.}{2005}]{casella2005ApJ...629..403C}
{Casella} P.,  {Belloni} T.,   {Stella} L.,  2005, \mn@doi [\apj]
  {10.1086/431174}, \href
  {https://ui.adsabs.harvard.edu/abs/2005ApJ...629..403C} {629, 403}

\bibitem[\protect\citeauthoryear{{Chakraborty}, {Navale}, {Ratheesh}  \&
  {Bhattacharyya}}{{Chakraborty} et~al.}{2020}]{Chakraborty2020MNRAS.498.5873C}
{Chakraborty} S.,  {Navale} N.,  {Ratheesh} A.,   {Bhattacharyya} S.,  2020,
  \mn@doi [\mnras] {10.1093/mnras/staa2711}, \href
  {https://ui.adsabs.harvard.edu/abs/2020MNRAS.498.5873C} {498, 5873}

\bibitem[\protect\citeauthoryear{{De Marco}, {Zdziarski}, {Ponti}, {Migliori},
  {Belloni}, {Segovia Otero}, {Dzie{\l}ak}  \& {Lai}}{{De Marco}
  et~al.}{2021}]{demarco2021arXiv210207811D}
{De Marco} B.,  {Zdziarski} A.~A.,  {Ponti} G.,  {Migliori} G.,  {Belloni}
  T.~M.,  {Segovia Otero} A.,  {Dzie{\l}ak} M.,   {Lai} E.~V.,  2021, arXiv
  e-prints, \href {https://ui.adsabs.harvard.edu/abs/2021arXiv210207811D} {p.
  arXiv:2102.07811}

\bibitem[\protect\citeauthoryear{{Done}, {Gierli{\'n}ski}  \& {Kubota}}{{Done}
  et~al.}{2007}]{Done2007A&ARv..15....1D}
{Done} C.,  {Gierli{\'n}ski} M.,   {Kubota} A.,  2007, \mn@doi [\aapr]
  {10.1007/s00159-007-0006-1}, \href
  {http://adsabs.harvard.edu/abs/2007A%26ARv..15....1D} {15, 1}

\bibitem[\protect\citeauthoryear{{Dzie{\l}ak}, {De Marco}  \&
  {Zdziarski}}{{Dzie{\l}ak} et~al.}{2021}]{dzielak2021arXiv210211635D}
{Dzie{\l}ak} M.~A.,  {De Marco} B.,   {Zdziarski} A.~A.,  2021, arXiv e-prints,
  \href {https://ui.adsabs.harvard.edu/abs/2021arXiv210211635D} {p.
  arXiv:2102.11635}

\bibitem[\protect\citeauthoryear{{Fabian} et~al.,}{{Fabian}
  et~al.}{2020}]{Fabian2020MNRAS.493.5389F}
{Fabian} A.~C.,  et~al., 2020, \mn@doi [\mnras] {10.1093/mnras/staa564}, \href
  {https://ui.adsabs.harvard.edu/abs/2020MNRAS.493.5389F} {493, 5389}

\bibitem[\protect\citeauthoryear{{Foreman-Mackey}, {Hogg}, {Lang}  \&
  {Goodman}}{{Foreman-Mackey} et~al.}{2013}]{emcee}
{Foreman-Mackey} D.,  {Hogg} D.~W.,  {Lang} D.,   {Goodman} J.,  2013, \mn@doi
  [PASP] {10.1086/670067}, 125, 306

\bibitem[\protect\citeauthoryear{{Gandhi}, {Rao}, {Johnson}, {Paice}  \&
  {Maccarone}}{{Gandhi} et~al.}{2019}]{2019MNRAS.485.2642G}
{Gandhi} P.,  {Rao} A.,  {Johnson} M. A.~C.,  {Paice} J.~A.,   {Maccarone}
  T.~J.,  2019, \mn@doi [\mnras] {10.1093/mnras/stz438}, \href
  {https://ui.adsabs.harvard.edu/abs/2019MNRAS.485.2642G} {485, 2642}

\bibitem[\protect\citeauthoryear{{Gendreau} et~al.,}{{Gendreau}
  et~al.}{2016}]{Gendreau2016SPIE.9905E..1HG}
{Gendreau} K.~C.,  et~al., 2016, {The Neutron star Interior Composition
  Explorer (NICER): design and development}.
p. 99051H, \mn@doi{10.1117/12.2231304}

\bibitem[\protect\citeauthoryear{{Guan} et~al.,}{{Guan}
  et~al.}{2020}]{guan2020arXiv201212067G}
{Guan} J.,  et~al., 2020, arXiv e-prints, \href
  {https://ui.adsabs.harvard.edu/abs/2020arXiv201212067G} {p. arXiv:2012.12067}

\bibitem[\protect\citeauthoryear{{Homan} et~al.,}{{Homan}
  et~al.}{2020}]{Homan2020ApJ...891L..29H}
{Homan} J.,  et~al., 2020, \mn@doi [\apjl] {10.3847/2041-8213/ab7932}, \href
  {https://ui.adsabs.harvard.edu/abs/2020ApJ...891L..29H} {891, L29}

\bibitem[\protect\citeauthoryear{{Ingram} \& {Motta}}{{Ingram} \&
  {Motta}}{2020}]{Ingram2020arXiv200108758I}
{Ingram} A.,  {Motta} S.,  2020, arXiv e-prints, \href
  {https://ui.adsabs.harvard.edu/abs/2020arXiv200108758I} {p. arXiv:2001.08758}

\bibitem[\protect\citeauthoryear{{Kara} et~al.,}{{Kara}
  et~al.}{2019}]{kara2019Natur.565..198K}
{Kara} E.,  et~al., 2019, \mn@doi [\nat] {10.1038/s41586-018-0803-x}, \href
  {https://ui.adsabs.harvard.edu/abs/2019Natur.565..198K} {565, 198}

\bibitem[\protect\citeauthoryear{{Kawamuro} et~al.,}{{Kawamuro}
  et~al.}{2018}]{2018ATel11399....1K}
{Kawamuro} T.,  et~al., 2018, The Astronomer's Telegram, \href
  {https://ui.adsabs.harvard.edu/abs/2018ATel11399....1K} {11399, 1}

\bibitem[\protect\citeauthoryear{{Kennea}, {Marshall}, {Page}, {Palmer},
  {Siegel}  \& {Neil Gehrels Swift Observatory Team}}{{Kennea}
  et~al.}{2018}]{2018ATel11403....1K}
{Kennea} J.~A.,  {Marshall} F.~E.,  {Page} K.~L.,  {Palmer} D.~M.,  {Siegel}
  M.~H.,   {Neil Gehrels Swift Observatory Team} 2018, The Astronomer's
  Telegram, \href {https://ui.adsabs.harvard.edu/abs/2018ATel11403....1K}
  {11403, 1}

\bibitem[\protect\citeauthoryear{{Leahy}, {Darbro}, {Elsner}, {Weisskopf},
  {Sutherland}, {Kahn}  \& {Grindlay}}{{Leahy}
  et~al.}{1983}]{Leahy1983ApJ...266..160L}
{Leahy} D.~A.,  {Darbro} W.,  {Elsner} R.~F.,  {Weisskopf} M.~C.,  {Sutherland}
  P.~G.,  {Kahn} S.,   {Grindlay} J.~E.,  1983, \mn@doi [\apj]
  {10.1086/160766}, \href
  {https://ui.adsabs.harvard.edu/abs/1983ApJ...266..160L} {266, 160}

\bibitem[\protect\citeauthoryear{{McKinney}, {Tchekhovskoy}  \&
  {Blandford}}{{McKinney} et~al.}{2012}]{mckinney2012MNRAS.423.3083M}
{McKinney} J.~C.,  {Tchekhovskoy} A.,   {Blandford} R.~D.,  2012, \mn@doi
  [\mnras] {10.1111/j.1365-2966.2012.21074.x}, \href
  {https://ui.adsabs.harvard.edu/abs/2012MNRAS.423.3083M} {423, 3083}

\bibitem[\protect\citeauthoryear{{Miller-Jones} et~al.,}{{Miller-Jones}
  et~al.}{2019}]{MJ2019Natur.569..374M}
{Miller-Jones} J. C.~A.,  et~al., 2019, \mn@doi [\nat]
  {10.1038/s41586-019-1152-0}, \href
  {https://ui.adsabs.harvard.edu/abs/2019Natur.569..374M} {569, 374}

\bibitem[\protect\citeauthoryear{{Miller}, {Reynolds}, {Fabian}, {Miniutti}  \&
  {Gallo}}{{Miller} et~al.}{2009}]{Miller2009ApJ...697..900M}
{Miller} J.~M.,  {Reynolds} C.~S.,  {Fabian} A.~C.,  {Miniutti} G.,   {Gallo}
  L.~C.,  2009, \mn@doi [\apj] {10.1088/0004-637X/697/1/900}, \href
  {https://ui.adsabs.harvard.edu/abs/2009ApJ...697..900M} {697, 900}

\bibitem[\protect\citeauthoryear{{Motta}, {Belloni}, {Stella},
  {Mu{\~n}oz-Darias}  \& {Fender}}{{Motta}
  et~al.}{2014a}]{Motta2014MNRAS.437.2554M}
{Motta} S.~E.,  {Belloni} T.~M.,  {Stella} L.,  {Mu{\~n}oz-Darias} T.,
  {Fender} R.,  2014a, \mn@doi [\mnras] {10.1093/mnras/stt2068}, \href
  {https://ui.adsabs.harvard.edu/abs/2014MNRAS.437.2554M} {437, 2554}

\bibitem[\protect\citeauthoryear{{Motta}, {Munoz-Darias}, {Sanna}, {Fender},
  {Belloni}  \& {Stella}}{{Motta} et~al.}{2014b}]{Motta2014MNRAS.439L..65M}
{Motta} S.~E.,  {Munoz-Darias} T.,  {Sanna} A.,  {Fender} R.,  {Belloni} T.,
  {Stella} L.,  2014b, \mn@doi [\mnras] {10.1093/mnrasl/slt181}, \href
  {https://ui.adsabs.harvard.edu/abs/2014MNRAS.439L..65M} {439, L65}

\bibitem[\protect\citeauthoryear{{Motta}, {Franchini}, {Lodato}  \&
  {Mastroserio}}{{Motta} et~al.}{2018}]{Motta2018MNRAS.473..431M}
{Motta} S.~E.,  {Franchini} A.,  {Lodato} G.,   {Mastroserio} G.,  2018,
  \mn@doi [\mnras] {10.1093/mnras/stx2358}, \href
  {https://ui.adsabs.harvard.edu/abs/2018MNRAS.473..431M} {473, 431}

\bibitem[\protect\citeauthoryear{{Mudambi}, {Maqbool}, {Misra}, {Hebbar},
  {Yadav}, {Gudennavar}  \& {S.~G.}}{{Mudambi}
  et~al.}{2020}]{Mudambi2020ApJ...889L..17M}
{Mudambi} S.~P.,  {Maqbool} B.,  {Misra} R.,  {Hebbar} S.,  {Yadav} J.~S.,
  {Gudennavar} S.~B.,   {S.~G.} B.,  2020, \mn@doi [\apjl]
  {10.3847/2041-8213/ab66bc}, \href
  {https://ui.adsabs.harvard.edu/abs/2020ApJ...889L..17M} {889, L17}

\bibitem[\protect\citeauthoryear{{Nowak}, {Wagoner}, {Begelman}  \&
  {Lehr}}{{Nowak} et~al.}{1997}]{Nowak1997ApJ...477L..91N}
{Nowak} M.~A.,  {Wagoner} R.~V.,  {Begelman} M.~C.,   {Lehr} D.~E.,  1997,
  \mn@doi [\apjl] {10.1086/310534}, \href
  {https://ui.adsabs.harvard.edu/abs/1997ApJ...477L..91N} {477, L91}

\bibitem[\protect\citeauthoryear{{Psaltis}, {Belloni}  \& {van der
  Klis}}{{Psaltis} et~al.}{1999}]{PBK1999ApJ...520..262P}
{Psaltis} D.,  {Belloni} T.,   {van der Klis} M.,  1999, \mn@doi [\apj]
  {10.1086/307436}, \href
  {https://ui.adsabs.harvard.edu/abs/1999ApJ...520..262P} {520, 262}

\bibitem[\protect\citeauthoryear{{Remillard} \& {McClintock}}{{Remillard} \&
  {McClintock}}{2006}]{remillard2006ARA&A..44...49R}
{Remillard} R.~A.,  {McClintock} J.~E.,  2006, \mn@doi [\araa]
  {10.1146/annurev.astro.44.051905.092532}, \href
  {https://ui.adsabs.harvard.edu/abs/2006ARA&A..44...49R} {44, 49}

\bibitem[\protect\citeauthoryear{{Reynolds}}{{Reynolds}}{2020}]{SpinRev2020arXiv201108948R}
{Reynolds} C.~S.,  2020, arXiv e-prints, \href
  {https://ui.adsabs.harvard.edu/abs/2020arXiv201108948R} {p. arXiv:2011.08948}

\bibitem[\protect\citeauthoryear{{Reynolds} \& {Miller}}{{Reynolds} \&
  {Miller}}{2009}]{reynolds2009ApJ...692..869R}
{Reynolds} C.~S.,  {Miller} M.~C.,  2009, \mn@doi [\apj]
  {10.1088/0004-637X/692/1/869}, \href
  {https://ui.adsabs.harvard.edu/abs/2009ApJ...692..869R} {692, 869}

\bibitem[\protect\citeauthoryear{{Shidatsu} et~al.,}{{Shidatsu}
  et~al.}{2018}]{Shidatsu2018ApJ...868...54S}
{Shidatsu} M.,  et~al., 2018, \mn@doi [\apj] {10.3847/1538-4357/aae929}, \href
  {https://ui.adsabs.harvard.edu/abs/2018ApJ...868...54S} {868, 54}

\bibitem[\protect\citeauthoryear{{Shidatsu}, {Nakahira}, {Murata}, {Adachi},
  {Kawai}, {Ueda}  \& {Negoro}}{{Shidatsu}
  et~al.}{2019}]{Shidatsu2019ApJ...874..183S}
{Shidatsu} M.,  {Nakahira} S.,  {Murata} K.~L.,  {Adachi} R.,  {Kawai} N.,
  {Ueda} Y.,   {Negoro} H.,  2019, \mn@doi [\apj] {10.3847/1538-4357/ab09ff},
  \href {https://ui.adsabs.harvard.edu/abs/2019ApJ...874..183S} {874, 183}

\bibitem[\protect\citeauthoryear{{Stella} \& {Vietri}}{{Stella} \&
  {Vietri}}{1998}]{Stella1998ApJ...492L..59S}
{Stella} L.,  {Vietri} M.,  1998, \mn@doi [\apjl] {10.1086/311075}, \href
  {https://ui.adsabs.harvard.edu/abs/1998ApJ...492L..59S} {492, L59}

\bibitem[\protect\citeauthoryear{{Stella}, {Vietri}  \& {Morsink}}{{Stella}
  et~al.}{1999}]{Stella1999ApJ...524L..63S}
{Stella} L.,  {Vietri} M.,   {Morsink} S.~M.,  1999, \mn@doi [\apjl]
  {10.1086/312291}, \href
  {https://ui.adsabs.harvard.edu/abs/1999ApJ...524L..63S} {524, L63}

\bibitem[\protect\citeauthoryear{{Stiele} \& {Kong}}{{Stiele} \&
  {Kong}}{2020}]{Stiele2020ApJ...889..142S}
{Stiele} H.,  {Kong} A.~K.~H.,  2020, \mn@doi [\apj]
  {10.3847/1538-4357/ab64ef}, \href
  {https://ui.adsabs.harvard.edu/abs/2020ApJ...889..142S} {889, 142}

\bibitem[\protect\citeauthoryear{{Strohmayer}}{{Strohmayer}}{2001}]{strohmayer2001ApJ...552L..49S}
{Strohmayer} T.~E.,  2001, \mn@doi [\apjl] {10.1086/320258}, \href
  {https://ui.adsabs.harvard.edu/abs/2001ApJ...552L..49S} {552, L49}

\bibitem[\protect\citeauthoryear{{Torres}, {Casares}, {Jim{\'e}nez-Ibarra},
  {Mu{\~n}oz-Darias}, {Armas Padilla}, {Jonker}  \& {Heida}}{{Torres}
  et~al.}{2019}]{Torres2019ApJ...882L..21T}
{Torres} M.~A.~P.,  {Casares} J.,  {Jim{\'e}nez-Ibarra} F.,  {Mu{\~n}oz-Darias}
  T.,  {Armas Padilla} M.,  {Jonker} P.~G.,   {Heida} M.,  2019, \mn@doi
  [\apjl] {10.3847/2041-8213/ab39df}, \href
  {https://ui.adsabs.harvard.edu/abs/2019ApJ...882L..21T} {882, L21}

\bibitem[\protect\citeauthoryear{{Torres}, {Casares}, {Jim{\'e}nez-Ibarra},
  {{\'A}lvarez-Hern{\'a}ndez}, {Mu{\~n}oz-Darias}, {Armas Padilla}, {Jonker}
  \& {Heida}}{{Torres} et~al.}{2020}]{Torres2020ApJ...893L..37T}
{Torres} M.~A.~P.,  {Casares} J.,  {Jim{\'e}nez-Ibarra} F.,
  {{\'A}lvarez-Hern{\'a}ndez} A.,  {Mu{\~n}oz-Darias} T.,  {Armas Padilla} M.,
  {Jonker} P.~G.,   {Heida} M.,  2020, \mn@doi [\apjl]
  {10.3847/2041-8213/ab863a}, \href
  {https://ui.adsabs.harvard.edu/abs/2020ApJ...893L..37T} {893, L37}

\bibitem[\protect\citeauthoryear{{Zdziarski}, {Dzie{\l}ak}, {De Marco},
  {Szanecki}  \& {Nied{\'z}wiecki}}{{Zdziarski}
  et~al.}{2021}]{zdz2021ApJ...909L...9Z}
{Zdziarski} A.~A.,  {Dzie{\l}ak} M.~A.,  {De Marco} B.,  {Szanecki} M.,
  {Nied{\'z}wiecki} A.,  2021, \mn@doi [\apjl] {10.3847/2041-8213/abe7ef},
  \href {https://ui.adsabs.harvard.edu/abs/2021ApJ...909L...9Z} {909, L9}

\makeatother
\end{thebibliography}




\appendix

%


\bsp	
\label{lastpage}
\end{document}